\documentclass{pasj00}
\draft

\begin{document}
\SetRunningHead{Hiraga et al.}{Search Sc-K line emission with Suzaku}
\Received{2008/09/16}
\Accepted{2008/09/00}

\title{Search for Sc-K line emission from RX~J0852.0--4622 Supernova remnant with Suzaku}

\author{Junko S. \textsc{hiraga},\altaffilmark{1}
Yusuke \textsc{kobayashi}, \altaffilmark{2}
Toru \textsc{tamagawa}, \altaffilmark{1,3}
Asami \textsc{hayato}, \altaffilmark{1,3}
Aya \textsc{bamba}, \altaffilmark{2}
Yukikatsu \textsc{Terada}, \altaffilmark{4}
Robert~\textsc{petre},~\altaffilmark{5}
Hideaki \textsc{Katagiri}, \altaffilmark{6}
\and Hiroshi \textsc{Tsunemi}, \altaffilmark{7}}

\altaffiltext{1}{RIKEN, 2-1 Hirosawa, Wako, Saitama 3510198}
\email{jhiraga@crab.riken.jp}
\altaffiltext{2}{Institute of Space and Astronautical Science, Japan Aerospace Exploration Agency, 3-1-1 Yoshinodai Sagamihara, Kanagawa 2298510}
\altaffiltext{3}{Department of Physics, Tokyo University of Science, 1-3 Kagurazaka, Shinjyuku-ku, Tokyo 1628601}
\altaffiltext{4}{Department of Physics, Saitama University, Shimo-Okubo 255, Sakura, Saitama 3388570, Japan}
\altaffiltext{5}{Astrophysics Science Division, NASA\/Goddard Space Flight Center, Code 662, Greenbelt, MD 20771, USA}
\altaffiltext{6}{Department of Physical Science, Graduate School of Science, Hiroshima University 1-3-1 Kagamiyama, Higashi-hiroshima, 7398526 Japan}
\altaffiltext{7}{Graduate School of Science, Earth and Space Science,Osaka University, Machikaneyama, 1-1Toyonaka, Osaka, 5600043,  Japan}


%


\maketitle

\begin{abstract}
We searched for evidence of line emission around 4\,keV from the northwestern rim of the supernova remnant RX~J0852.0-4622 using Suzaku XIS data.  Several papers have reported the detection of an emission line around 4.1\,keV from this region of the sky. This line would arise from K-band fluorescence by  $^{44}$Sc, the immediate decay product of $^{44}$Ti. 
We performed spectral analysis for the entire portion of the NW rim of the remnant within the XIS field of view, as well as various regions corresponding to regions of published claims of line emission. We found no line emission around 4.1\,keV anywhere, and are able to set a restrictive upper limit to the X-ray flux:   1.1$\times$10$^{-6}$ s$^{-1}$ cm$^{-2}$ for the entire field. For every region, our flux upper limit falls below that of the previously claimed detection.  Therefore, we conclude that, to date, no definite X-ray line feature from Sc-K emission has been detected in the NW rim of RX~J0852.0-4622. Our negative-detection supports the recent claim that RX~J0852-4622 is neither young~(1700--4000 yr) nor nearby~($\sim$750 pc). 

\end{abstract}

\section{Introduction}
Supernovae are believed to be an agent for producing heavy elements and distributing them to the Galaxy.  The theory of supernova explosions predicting nucleosynthesis yields has been developed over the past 50 years. However, there is no observational evidence which allows us to carry out quantitative comparison with these models, and the explosions themselves are still not understood.

$^{44}$Ti\,\,is a short-lived radioisotope with a half-life of about 60 yrs. It is thought to be produced by explosive nucleosynthesis in SNe and to be the source of the stable $^{44}$Ca\,\, in our Galaxy. The abundunce of this species strongly depends on the explosion details, mainly on the so-called ``mass-cut"  in core-collapse SNe, the energy of the explosion and explosion asymmetries. Thus,  the decay-chain of $^{44}$Ti\,\, offers a unique window into the study of the supernova explosion mechanism~(see review \citet{diehl98}).  The radioactive decay chain $^{44}$Ti$\to$$^{44}$Sc$\to$$^{44}$Ca\,\, produces three gamma-ray lines at 67.9\,keV, 78.4\,keV and 1157\,keV with similar branching ratios. Several gamma-ray space missions, including COMPTEL, BeppoSAX and INTEGRAL, have sought these lines in young supernova remnants. The discovery of the $^{44}$Ti\,\,1157\,keV line emission from the young famous Galactic supernova remnant Cas A with COMPTEL(\cite{iyudin94}) was the first direct proof that this isotope is really produced in SNe. 
Subsequent detection of line emission around 70\,keV by BeppoSAX and INTEGRAL has strengthened this result(\cite{vink01, renaud06}).  Today, Cas A remains the only SNR from which  $^{44}$Ti\,\, lines have been clearly detected. 
GRO J0852-4642 has been also reported as a $^{44}$Ti\,\, emitter with a flux of 3.8$\pm$0.7 $\times$10$^{-5}$cm$^{-2}$s$^{-1}$ based on six years of COMPTEL data~(\cite{iyudin98}).  \citet{schon00} examined the robustness of the $^{44}$Ti\,  line detection, applying different background modeling and event selection criteria to suppress a large part of the background. They found a significant variation for the line flux ranging from 2$\sigma$ to 4$\sigma$ depending on analysis method, whereas the Cas A significance hardly varies~(4$\sigma$ or more).  The authors, therefore, conclude that the $^{44}$Ti\, line detection for GRO J0852--4642 is marginal even though it is  the second brightest feature in COMPTEL's survey in the 1157\,keV band.  For the 67.9\,keV and 78.4\,keV lines from GRO J0852--4642, \cite{keinlin04} obtained a flux upper limit of 1.1$\times$ 10$^{-4}$ cm$^{-2}$s$^{-1}$ with the INTEGRAL/SPI. The large value is mainly due to systematic uncertainty.

$^{44}$Ti\,\, is an unstable, proton-rich isotope among the many new nuclei synthesized in supernova explosions. Many interesting proton-rich nucleosynthesis products decay by electron capture, leaving primarily K-shell vacancies in the daughter atoms. The adjustment of atomic electrons in response to the vacancies causes the emission of characteristic X-rays (\cite{leising01}). For the case of $^{44}$Ti\,\, decay, it produces Sc-K$\alpha$(4.086, 4.091\, keV) line emission.  Currently, and for the near future, X-ray observatories have much larger effective area, better imaging capability, lower background level and better energy resolution than Gamma-ray observatories; i.e., it is generally suitable for line detection in X-ray band rather than Gamma-ray one. Therefore searching for the X-ray  lines offers an attractive alternative method of direct measurement of the mass of $^{44}$Ti\,\, ejected and furthermore the location of the $^{44}$Ti\,\, within the remnant. Although the most promising object where substantial $^{44}$Ti\,\, should exist is Cas A,  its large continuum flux prevents us clearly detecting this X-ray line. On the other hand, detection of Sc-K emission from the northwestern (NW) region of SNR RXJ0852.0--4622(Vela jr.) have been reported in several studies.  RXJ0852.0--4622 was recently discovered  in the southeastern corner of the Vela SNR with a large apparent radius of about 60\arcmin~(\cite{acb98}) and is coincident with GRO J0852.0--4622. \cite{tsunemi00} reported on the presence of line emission around 4.1\.keV using an ASCA SIS observation, and concluded that there is substantial Ca produced from $^{44}$Ti\,.  On the other hand,  \cite{slane01}, also using ASCA data, found no evidence for a line, and gave a 1$\sigma$ upper limit of Sc-K line flux as 4.4$\times$10$^{-6}$ cm$^{-2}$s$^{-1}$.  \citet{iyudin05} reported on the detection of Sc-K line emission from various regions in the remnant using XMM-Newton data. \cite{bamba05} also reported on a possible excess around 4.1\,keV from a Chandra observation. 

In this paper, we take advantage of the large effective area around 4.1\,keV of Suzaku to make the most sensitive search to date for the Sc-K line emission. Spectra extracted from corresponding regions to those in previously published papers were examined for the existence of an emission line feature.  Other X-ray properties of the remnant such as nonthermal emission and thin-thermal plasma will be discussed in  separate papers.

\section{Observation and Data Reduction}
\begin{figure}
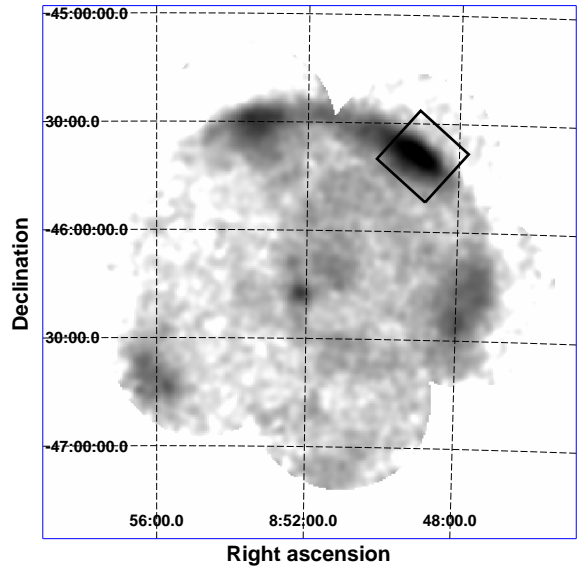

  \begin{center}
    \FigureFile(100mm,100mm){fig1_suzakufov.ps}
  \end{center}
  \caption{ASCA observation of RX~J0852.0-4622.  Suzaku FOV is overlayed as a solid black square.}\label{fig1:asca}
\end{figure}

The NW part  of RX~J0852.0-4622  was observed using the Japanese-US X-ray astronomy satellite, Suzaku~(\cite{mitsuda07}) on 2005 December 19 during the Performance Verification~(PV) phase. This is the brightest portion in the remnant, with a large apparent radius of about 60\arcmin.  The field of view~(FOV) of our Suzaku observations overlaid on ASCA image of the whole remnant~(\cite{tsunemi00}) is shown  in figure\,\ref{fig1:asca}. We also observed a nearby sky region with the same Galactic latitude for background determination. Suzaku carries four X-ray Imaging Spectrometers~(XIS;\cite{koyama07}) and the non-imaging Hard X-ray Detector~(HXD;\cite{takahashi07}). Each XIS consists of a CCD detector at the focal plane of a dedicated thin-foil X-ray telescope~(XRT;\cite{serlemitsos07}) with 18\arcmin$\times$18\arcmin field of view. One of the XIS detectors~(XIS\,1) is back-illuminated~(BI) and the other three (XIS0, XIS2 and XIS3) are front-illuminated (FI).   The advantages of the former are significantly superior sensitivity in the 0.2--1.0\,keV band with moderate energy resolution while the latter has high detection efficiency and low background level in the energy band above $\sim$5\,keV. 

The XIS was operated in the normal full-frame clocking mode without spaced-raw charge injection~(SCI) during both observations and the data format was 3$\times$3 or 5$\times$5.  The data were processed and cleaned using version 2.0.6.13 of the standard Suzaku pipeline software. The CALDB version of the products is hxd-20070710, xis-20070731, and xrt-20070622. We used HEADAS software 6.3.2 and XSPEC version 11.3.2 for the data reduction and analysis. The net exposure times are approximately 175\,ksec and 59\,ksec for source and background respectively.  The observations are summarized in table\ref{tab1:log}. Since the characteristics of the three FI sensors are almost identical and well calibrated to each other, we combined the data from them~(hereafter XIS\,023) and show the average spectrum.
In this paper, we concentrate on searching for Sc-K line emission. For that purpose, we present an analysis of only the XIS\,023 data. XIS\,023 has the largest effective area at 4 keV among current X-ray observatories, $\sim$900\,cm$^{2}$ .  

\section{Analysis and Results}
\begin{figure}
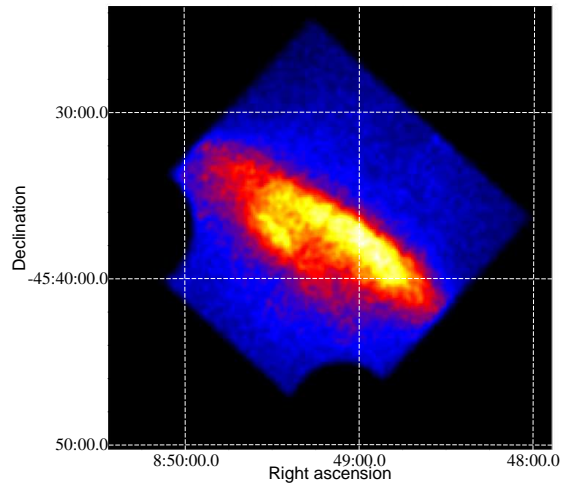

  \begin{center}
      \FigureFile(110mm,110mm){fig2_xis_image.ps}
  \end{center}
  \caption{Suzaku XIS0 Image of RX~J0852.0--4622 NW rim. 2-8\,keV photons are used excluding calibration source on two corners.}\label{fig2:xis}
\end{figure}

Figure\,\ref{fig2:xis} shows the X-ray image obtained by XIS\,023 in the energy range of 2.0--8.0\,keV.  Two bright rims, an outer and an inner one, are seen as revealed by previous observations using XMM-Newton and Chandra (\cite{iyudin05,bamba05}).  In the XIS\,023 image, they are not distinctly separated. 

\subsection{Spectrum from all of rim regioin}\label{rimall}
\begin{figure}
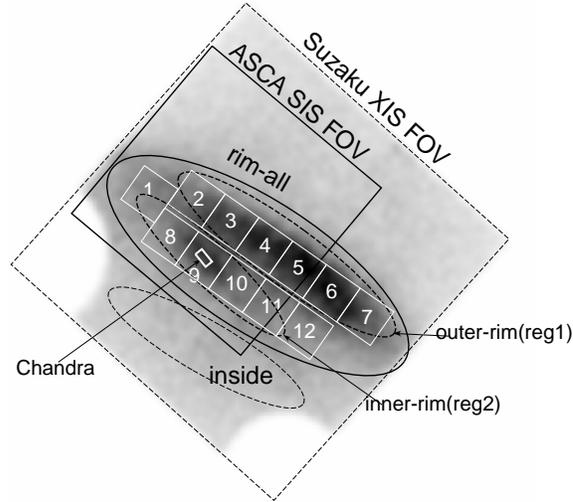

  \begin{center}
 \hspace*{-20mm}
   \FigureFile(120mm,150mm){fig3_specreg.ps}
  \end{center}
  \caption{Various regions where we extracted spectra are shown on gray scale Suzaku image, which is the  same as figure.\ref{fig2:xis}. Labels of respective regions are described in text.}\label{fig3:spec_reg}.  
\end{figure}

We extracted a representative XIS\,023 spectrum from an elliptical region including both outer and inner rims as identified by ``rim-all"  in figure\,\ref{fig3:spec_reg}. This region includes the site where several papers reported the detection of line emission around 4.1\.keV~(\cite{tsunemi00,slane01,bamba05, iyudin05}).

The responses of the XRT and XIS were calculated using the ``xissimarfgen"  version 2006-11-26 ancillary response file~(ARF) generator (\cite{ishisaki07}) and the ``xisrmfgen" version 2006-11-26 response matrix file~(RMF) generator. The energy resolution of this data set was $\sim$110\,eV~(FWHM) at 4\,keV~(\cite{koyama07}). A slight degradation of the energy resolution was included in the RMF. A decrease of the low-energy transmission of the XIS optical blocking filter~(OBF) was included in the ARF, but this effect has no influence in the 2.0--8.0\,keV energy band  being used in this paper.  The ARF response was calculated assuming that photons come from only the rim-all region ~(shown in figure\,\ref{fig3:spec_reg}) with a flat surface brightness profile. We checked that contamination from source flux of outside of rim-all is negligible using the region interior to the rim~(shown in figure\,\ref{fig3:spec_reg}) which is about 1/4 dimmer than rim-all region.

A background spectrum was extracted from the offset data using the same detector area as the rim-all region. There is no significant signal beyond the background level above 8.0\,keV. Additionally, in order to avoid uncertainty about the thermal emission which could have contributions  from both the Vela SNR and Vela Jr. below 2\,keV, we used photons in the energy range in 2.0--8.0\,keV for spectral fitting.

\begin{figure}
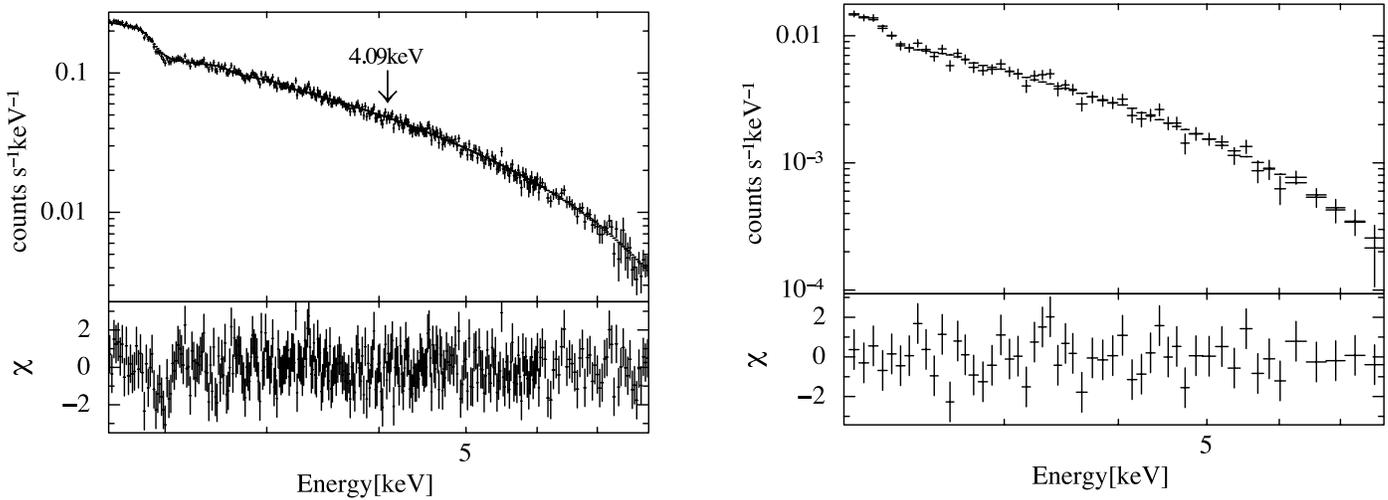

  \begin{center}
\hspace*{-24mm}
 @\FigureFile(45mm,100mm){fig4a_rimall_1.eps}
\hspace*{50mm}
       \FigureFile(45mm,50mm){fig4b_reg9_1.eps}
  \end{center}
  \caption{Suzaku spectrum from rim-all region of RX~J0852.0-4622 in left panel. Right panel shows spectrum from a representative small square region (region ID=9).   No line emission is detected around 4keV}\label{fig4:spec}
\end{figure}

The resultant  XIS~023 spectrum is apparently dominated by nonthermal emission as shown in the left panel of figure \,\ref{fig4:spec}. It is well represented by a single power law spectrum with interstellar absorption. The derived photon index of $\Gamma$= 2.81$\pm$0.05 and absorbing column density of $N_{\rm{H}}$=(0.67$\pm$0.11)$\times$10$^{22}$cm$^{-2}$ with $\chi^2/$d.o.f=331.4/324 are in good agreement with previous results by ASCA, XMM-Newton and Chandra (\cite{slane01, iyudin05, bamba05}).
There appears to be no significant emission feature around 4\,keV.  In order to search for evidence from Sc-K line emission, we introduced into the model an additional gaussian component  at the fixed energy of 4.09\,keV. XIS has the absolute energy error within $\pm$5\,eV~(\cite{koyama07}).  However, adding this component does not improve the fit at all, and gives an upper-bound for the line flux of 1.2$\times$10$^{-6}$ \,cm$^{-2}$s$^{-1}$\, (90\% confidence). Here, the width of channel bin is set to $\sim$15\,eV around 4.1\,keV. If left as a free parameter, the line centroid energy is unconstrained. We therefore performed the spectral fitting with a fixed energy.  

\subsection{Spatial Divided Spectral analysis}\label{2by2}
For more detailed study, we divided rim-all region into 12 square small cells indicated in figure\,\ref{fig3:spec_reg}. The size of one cell is 2\arcmin$\times$2\arcmin  \,which is comparable to the point spread function~(PSF) of the XRT. In analyzing the spectrum from  each cell, a corresponding background spectrum was extracted from the offset data. The responses of the XRT and XIS were calculated assuming the same region as for the rim-all analysis~(see \S\,\ref{rimall}) and normalized by the ratio between the cell size and the assumed source area~(that of ``rim-all" in this case).   All spectra were well fitted by a single power law with interstellar absorption. Figure\,\ref{fig4:spec}~(right) shows the resultant spectrum of region ID 9 as a representative spectrum. Here, the width of channel bin is set to $\sim$100\,eV around 4.1\,keV. No significant excess around 4.1\,keV is found in any cell, with upper limits of $F_{{\rm line}} \leq$(0.15--0.54)$\times$10$^{-6}$\,cm$^{-2}$s$^{-1}$\, depending on the surface brightness of the cells.

\subsection{Comparison with Previous Results}
Several papers  report a line emission feature around 4.1\,keV from this sky region. In order to directly compare our results  to previous work, we extracted spectra from regions corresponding to those used in the previous papers.  
All regions are shown in figure\,\ref{fig3:spec_reg}. The thin solid black square indicates the ASCA region used by \cite{slane01}; the small solid white rectangular represents the  Chandra detection region \cite{bamba05}. The solid and dashed thin elliptical regions are defined as outer and inner rim, respectively.
We performed spectral analysis for all regions mentioned above using the same procedure as for the other regions~(see \S\ref{rimall} and \S\ref{2by2}).  There appears to be no line emission in any spectrum; each is well represented by a single power law and interstellar absorption model.

For the  ``ASCA"  region, \cite{tsunemi00} suggested a significant excess around 4.1\,keV in the spectrum from the ASCA SIS, which has 11\arcmin$\times$11\arcmin\, FOV. Using thermal fits, they included the additional line-like feature with the $\sim$99\% confidence level according to the $F$-test.  On the other hand, \cite{slane01} reported that the spectrum from one of two detectors, SIS0,  contains  a feature at $\sim$4\,keV but not  the other, SIS1, of the same region. They derived a 1$\sigma$ upper limit  of 4.4 $\times$ 10$^{-6}$cm$^{-2}$s$^{-1}$ for Sc-K emission from this region. Using the Suzaku XIS data, we performed a spectral analysis for the ``ASCA"  region.  We found no line-like feature anywhere and derive a 90\% upper limit of $F_{{\rm line}} \leq$1.1$\times$10$^{-6}$\,cm$^{-2}$s$^{-1}$\,.

Using XMM-Newton data, \cite{iyudin05} reported the detection of some emission line features in 4.2--4.5\,keV band from both bright filaments of the NW rim. These two regions are described as ``reg1" and ``reg2" in the paper by \cite{iyudin05} but there is no information about their precise location. We therefore defined two elliptical regions~(outer and inner rims shown in figure\,\ref{fig3:spec_reg}) as corresponding regions for  ``reg1" and ``reg2" for Suzaku data. 
Spectral analysis for these regions were performed. Again, no line-like feature was found.  The 90 \% confidence upper limit for the outer rim~(reg1) is $F_{{\rm line}} \leq$5.7$\times$10$^{-7}$\,cm$^{-2}$s$^{-1}$\, and the upper limit for the inner rim~(reg2) is $F_{{\rm line}} \leq$1.1$\times$10$^{-6}$\,cm$^{-2}$s$^{-1}$\, at a fixed X-ray energy of 4.09\,keV.   Remarkably, our derived upper limits are about 6 or 2 times those lower than values claimed by \cite{iyudin05}.   We also verified the line emission at higher X-ray energy of 4.24\,keV for the outer rim and 4.2\,keV for the inner rim as same as claims by \cite{iyudin05}. There are not any line-like features giving the 90\% confidence upper limit of 1.2$\times$10$^{-6}$\,cm$^{-2}$s$^{-1}$\, for the outer rim and 5.0$\times$10$^{-7}$\,cm$^{-2}$s$^{-1}$\, for the inner rim.

\cite{bamba05} reported that there is an excess around 4\,keV from a tiny region on the inner filament indicated as the  ``Chandra" region in figure\,\ref{fig3:spec_reg}.  They obtained an acceptable fit including an additional gaussian component as a narrow line model with a power law model, although the reduced $\chi^{2}$ does not improve significantly ($\chi^2$/d.o.f changes from 76.2/81 to 69.5/79).  The derived flux was 7.3$^{+5.1}_{-4.5}$$\times$10$^{-7}$\,cm$^{-2}$s$^{-1}$. In order to compare the Chandra result with our evaluation, we employed region-ID 9 to serve as the corresponding region due to blurred PSF of Suzaku. It is noticed that our resultant upper limit of line flux, $F_{{\rm line}} \leq$3.8$\times$10$^{-7}$\, cm$^{-2}$s$^{-1}$\, is about half the value of the derived Chandra flux.

We summarize our derived upper limits of line flux for all regions in table\,\ref{tab2:result} where previous results are also listed. Figure.\ref{fig5:comparison} also presents our Suzaku evaluation with each corresponding published claims for direct comparison. It is noticed that our resultant upper limits indicate remarkably low flux compared with those obtained in  previous works. 

\section{Discussion and Conclusion}
RX~J0852.0--4622 is a unique object from which the possible detection of Sc-K line emission~(around 4.1\,keV) has been claimed (\cite{tsunemi00, iyudin05, bamba05}).  In order to search for evidence of line emission around 4.1\,keV,  we performed spectral analysis for various regions in the NW rim of RX~J0852.0--4622 with the Suzaku observatory.  No line-like features were found from any region, including the whole rim region contained in the Suzaku field of view, small subdivisions of the region, and regions from which detections have been claimed using ASCA, XMM-Newton and Chandra observations. The Suzaku line flux upper limits are 2--6 times lower than those of published claims for all regions. $^{44}$Ti would be ionized enough to shift the $^{44}$Sc-K lines to higher energy in such a shell of SNR. The featureless spectrum of Suzaku  imply that the similar results would be yielded even in higher energy.   Then we conclude that, to date, no credible X-ray line from Sc-K has been detected in the NW of RX~J0852.0--4622. In this study, we do not consider any other confusing lines as thermalized Sc-K line and  ionized Ca-K lines from thermal plasma of RX~J0852.0--4622 and/or Vela, which  overlays  in line of sight. 

Our upper limits presented here employ a 90\% confidence interval, determined using $\chi^{2}$ statistics.  For a single parameter of interest, this corresponds to an increase of $\chi^{2}$ by 2.7 from its minimum value.  Such a confidence limit can be compared directly with detection claims by XMM-Newton and Chandra, since they employed 90\% confidence intervals for their "detection" values, as is common for statistical error assessment in spectral fitting.  In light of the possible controversial difference between our result and others, we also examined the implications of using a more conservative confidence interval of 99.7\% ~(equivalent gaussian  width of 3$\sigma$).  According  to $\chi^2$-statistics,  a 3$\sigma$ confidence interval corresponds to $\Delta\chi^{2}$ of 9.0.   The 3$\sigma$ flux upper limit was derived to be 2.43$\times$10$^{-6}$\, cm$^{-2}$s$^{-1}$\, for the ``rim-all" region.  We also inferred $\Delta\chi^{2}$, for a flux of 0,  using the best fit parameter, where the $\chi^{2}$ value becomes the minimum,  and its 90\% error described in  \cite{iyudin05} and \cite{bamba05}~(see reg1, reg2 and chandra in Table.2).  In this regards, it is assumed to be the parabolic distribution near the $\chi^{2}$ minimum for easy comparison. The resultant $\Delta\chi^{2}$s were 5.0, 6.9 and 7.1 for ``reg1", ``reg2" and ``chandra" region, respectively. It is noticed all of ``detection" reports show low $\Delta\chi^{2}$ at 0 flux indicating less than 3$\sigma$ significance in the $\chi^{2}$ statistics. 

\begin{figure}
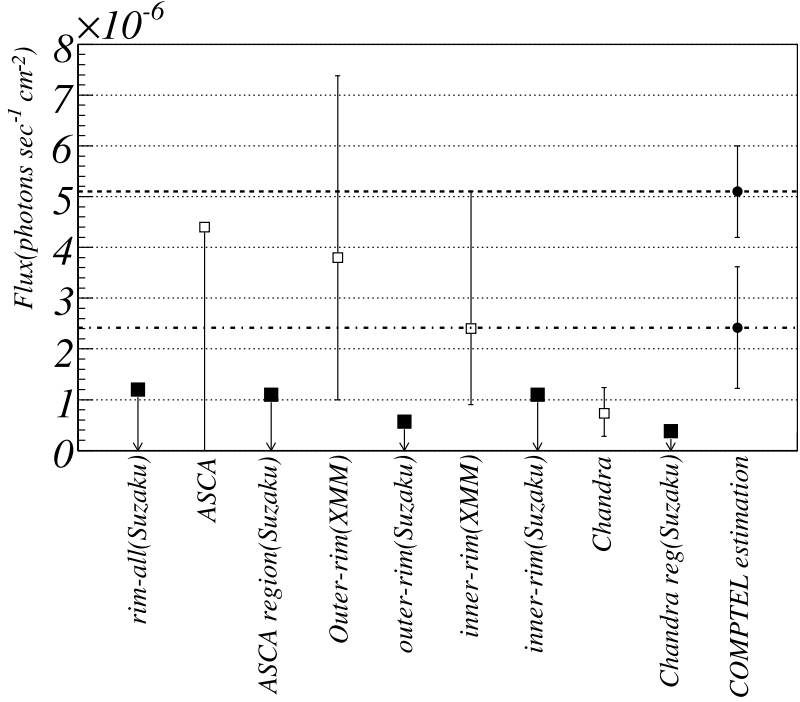

  \begin{center}
   \FigureFile(120mm,180mm){fig5_comparison.eps}
  \end{center}
  \caption{Direct comparison of line fluxes around 4.1\,keV between Suzaku XIS 90\% upper limits (black filled square) and previous claims of line detections~(white square). Estimated X-ray flux by using gamma-ray flux reported by \citet{iyudin98} and \citet{schon00} are shown as black filled circles.}\label{fig5:comparison}
\end{figure}

We investigate the consistency with observed gamma-ray flux, keeping in mind that the gamma-ray detection by COMPTEL is now considered marginal~(\cite{iyudin98, schon00}).  Expected X-ray flux of line emission, $F_{X}$,   can be estimated from the gamma-ray flux~($F_{\gamma}$) using the expression:
\begin{eqnarray}\label{eq1:flux}
F_{X} & = &\frac{F_{\gamma} g_{X} I_{X} f_{X}}{I_{\gamma} f_{\gamma}} 
\end{eqnarray}
,where $g_{X}$ is the K-shell electron capture fraction among the total number of decays, $I_{X}$ is the fluorescence yield of K$\alpha$ X-ray emission,  $I_{\gamma}$ is the absolute intensity of the flux per decay of the parent nucleus, and $f_{X}$ and  $f_{\gamma}$ are the escape fractions of the X- and  $\gamma$- photons, respectively, both of which are definitely equal to 1 for RX~J0852.0--4622.
Applying $g_{X}\sim$8/9, $I_{\gamma}$(1157\,keV)=0.999 and  $I_{X}$(Sc-K$\alpha$)$=$0.17, we can roughly estimate expected the $F_{X}$ from $F_{\gamma}$.
The decay of $F_{\gamma}$ due to about 10 years interval between gamma-ray observation and X-ray observation was considered.
The flux is estimated to be  5.1$\times$10$^{-6}$cm$^{-2}$s$^{-1}$ in the case of $F_{\gamma}=$3.8$\pm$0.7$\times$10$^{-6}$cm$^{-2}$s$^{-1}$(\cite{iyudin98}), indicated by dashed line in the fig.\ref{fig5:comparison}, and  2.4$\times$10$^{-6}$cm$^{-2}$s$^{-1}$ in the case of the lowest significance of gamma-ray source with $F_{\gamma}=$1.8$\pm$0.8$\times$10$^{-6}$cm$^{-2}$s$^{-1}$(see fig.4 in \cite{schon00}), shown as dotted and dashed line  in fig.\ref{fig5:comparison}.  It is found that the upper limits given by the Suzaku observation require lower flux that the lower value of the predicted X-ray flux. Our Suzaku result of no detection of Sc-K line emission is based on only the NW rim observation.  According to model calculation of nucleosynthesis in the supernova explosion, $^{44}$Ti is produced close to so-called ``mass-cut"  which is the innermost radius of ejected matter. It is possible that the line emission could appear from the interior of the remnant, and further observations and study are required. 

\citet{katsuda08} recently reported a slow X-ray expansion rate for this remnant of 0.23\% $\pm$ 0.006, based on a proper motion study using XMM-Newton observations taken over a span of 6.5\,yr. They estimated a remnant age of 1700-4000\,yr, depending on its evolutionary stage. Furthermore they estimated a distance $\sim$750\,pc, and assuming a high shock velocity of $\sim$3000\,km\,s$^{-1}$.  This result raises serious doubts about  the argument that RX~J0852.0--4622 is a young~($\sim$680 yr) and nearby~($\sim$200pc) supernova remnant (\cite{acb98}).   \citet{slane01} also suggested a larger distance of 1--2\,kpc based on a larger column density for this remnant than that for the Vela SNR from their ASCA analysis.  At this larger distance and with an older remnant age,  our negative-detection of Sc-K line emission becomes understandable.



\begin{table*}
  \caption{Suzaku Observation log of NW rim on RX J0852.0-4622 and its background region}\label{tab1:log}
  \begin{center}
    \begin{tabular}{lcccc}
      \hline
      Target  & Observatino ID & Coordinate & Observation start & Exposure(XIS) \\%
       &  & RA, Dec & (UT) & (ksec)\\
       \hline
     RXJ0852-4622 NW  &  500010010& \timeform{08h48m58s.01},\timeform{ -45D39'02''.9} & 2005-12-19&175 \\
     RXJ0852-4622 NW offset &  500010020& \timeform{09h00m17s.45},\timeform{ -47D56'39''.1} & 2005-12-23&59 \\
\hline
    \end{tabular}
  \end{center}
\end{table*}

\begin{table*}
  \caption{Comparison of detection and upper limits of Sc-K line emission.}\label{tab2:result}
  \begin{center}
    \begin{tabular}{lcllll}
      \hline
      Region ID & area& observatory  & flux & Energy & reference \\%
        &(arcmin${^2}$)& & (ph cm$^{-2}$ s $^{-1}$) & (keV) & \\
        \hline
 rim-all &91 & Suzaku & $<$1.2$\times$10$^{-6}$\footnotemark[$*$] & 4.09(fixed) & this work\\
        \hline
	 ASCA & 11$\times$11& ASCA & $<$4.4$\times$10$^{-6}$ \footnotemark[$\dagger$]& $\sim$4& \citet{slane01} \\
	  	& &Suzaku & $<$1.1$\times$10$^{-6}$\footnotemark[$*$]  & 4.09(fixed) & this work\\ 
\hline
  outer-rim(reg1) &31 & XMM-Newton  & 3.8(1.0-7.4)$\times$10$^{-6}$ \footnotemark[$\ddagger$] & 4.24(4.1-4.42) & \citet{iyudin05}\\
        & &Suzaku & $<$5.7$\times$10$^{-7}$\footnotemark[$*$]  & 4.09(fixed) & this work\\
         & &Suzaku & $<$1.0$\times$10$^{-6}$\footnotemark[$*$]  & 4.24(fixed) & this work\\
        \hline
       inner-rim(reg2) &17    & XMM-Newton & 2.4(0.9-5.1)$\times$10$^{-6}$\footnotemark[$\ddagger$]  & 4.2(4.0-4.2) & \citet{iyudin05}\\
        & &Suzaku & $<$1.1$\times$10$^{-6}$ \footnotemark[$*$] & 4.09(fixed) & this work\\
       & &Suzaku & $<$5.0$\times$10$^{-7}$\footnotemark[$*$]  & 4.2(fixed) & this work\\

\hline
       Chandra  &0.34  &Chandra& 7.3(2.8-12.4)$\times$10$^{-7}$ \footnotemark[$\ddagger$] & 4.11(3.9-4.42) & \citet{bamba05}\\
&4 & Suzaku(ID 9) & $<$3.8$\times$10$^{-7}$\footnotemark[$*$]  & 4.09(fixed) & this work\\	
 \hline
       \multicolumn{4}{@{}l@{}}{\hbox to 0pt{\parbox{85mm}{\footnotesize
 \par\noindent
 \footnotemark[$*$] 90\% confidence upper limit
 \par\noindent
\footnotemark[$\dagger$] 1 $\sigma$ confidence upper limit
\par\noindent
 \footnotemark[$\ddagger$] 90\% confidence error
 }\hss}}
    \end{tabular}
  \end{center}
\end{table*}

\end{document}